\documentclass[conference, 10pt, a4paper]{IEEEtran}
\IEEEoverridecommandlockouts
\usepackage{cite}
\usepackage{amsmath,amssymb,amsfonts}
\usepackage{algorithmic}
\usepackage{subfig}
\usepackage{graphicx}
\usepackage{textcomp}
\usepackage{xcolor}
\usepackage{booktabs}
\usepackage{standalone}
\usepackage{tikz}
\usepackage{booktabs}
\usetikzlibrary{calc, trees, positioning, arrows, shapes, shapes.geometric, fit, backgrounds, patterns}
\usepackage[capitalize]{cleveref}
\def\BibTeX{{\rm B\kern-.05em{\sc i\kern-.025em b}\kern-.08em
    T\kern-.1667em\lower.7ex\hbox{E}\kern-.125emX}}

\usepackage{pifont}
\newcommand{\cmark}{\ding{51}}
\newcommand{\xmark}{\ding{55}}

\begin{document}

\title{Perceptually-Weighted Video Quality Metric for Asymmetric Encoded Sports Videos}

\author{\IEEEauthorblockN{Anna Meyer\IEEEauthorrefmark{1}, Jonas Janzen\IEEEauthorrefmark{1}, Diwakara Reddy\IEEEauthorrefmark{1}, Alexander Kopte,\\
Simon Deniffel, Paul Wawerek-López, Marc Windsheimer, André Kaup}
\IEEEauthorblockA{\textit{Friedrich-Alexander-Universität Erlangen-Nürnberg, Erlangen, Germany}\\}
\IEEEauthorblockA{\IEEEauthorrefmark{1} \textit{These authors contributed equally}}}

\maketitle

\begin{abstract}
Objective video quality metrics commonly assume uniform spatial attention, an assumption that conflicts with the selective nature of human visual perception, particularly in sports videos. Here, allocating more bits for salient regions through semantic encoding can lead to significant bitrate savings. We present a Perceptually-Weighted Video Quality Metric (PW-VQM), a full-reference metric that accounts for the unequal perceptual importance of spatial regions and therefore targets quality evaluation for asymmetrically encoded content. SSIM maps computed in a multiscale wavelet domain are weighted by differentiating between foreground and background regions. Perceptually salient foreground regions are identified by combining open-vocabulary object detection with optical flow analysis, and are assigned higher weight during quality aggregation. Evaluated on sports video content, PW-VQM achieves a Spearman Rank Order Correlation Coefficient of 0.9511, outperforming established metrics including SSIM, VMAF, FUNQUE, and LPIPS. An ablation study confirms the individual contributions of the components of the perceptual weighting.
\end{abstract}


\section{Introduction}
Video Quality Assessment (VQA) plays a prominent role in the design, optimization, and evaluation of multimedia pipelines \cite{zhou2022}.
As the share of visual data in global internet data, including live sports streaming continues to grow, the ability to predict the perceptual quality experienced by end users and efficient transmission are becoming increasingly important. Semantic encoding that allocates more bits to important regions in a video offers a promising way to reduce bitrate if the viewing experience is not negatively affected. To optimize such asymmetric encoded video coding approaches, objective VQA metrics are required as proxies for costly and time-consuming subjective studies with humans in the loop.

Pixel-fidelity metrics, like Peak Signal-to-Noise Ratio~(PSNR) or Structural Similarity Index~(SSIM)~\cite{wang2004}, are well-established metrics to quantify visual quality. Metrics based on learned fusion strategies such as Video Multi-Method Assessment Fusion~(VMAF) or Fusion of Unified Quality Evaluators~(FUNQUE)~\cite{venkataramanan2022} as well as metrics incorporating perceptually motivated features like Learned Perceptual Image Patch Similarity~(LPIPS)~\cite{zhang2018} have been shown to better correlate with human judgment. Nevertheless, these metrics do not challenge the assumption that all spatial regions contribute equally to the overall perceived quality. However, for asymmetrically encoded content, viewers are more likely to notice distortions in critical areas surrounding semantically important regions. Here, encoding artifacts are more likely due to a change to a lower quality region. The QoMEX 2026 Grand Challenge on \textit{Video Quality Assessment for Asymmetric Encoded Videos} addresses the lack of VQA metrics that can accurately model perceived quality in asymmetric encoded videos by providing a sports video data set with human ratings. 

In this paper, we present our challenge solution, a Perceptually-Weighted Video Quality Metric (PW-VQM). Our metric uses a multiscale wavelet SSIM as backbone and introduces a novel perceptual weighting. First, we propose considering contrast sensitivity in the spatial as well as in the wavelet domain. Secondly, we identify perceptually important regions by a spatial weighting. Hereby, we combine object detection predictions that allow modeling asymmetric encoding with a motion-based importance rating that exploits domain-specific knowledge about sports content. 

\section{Perceptually-Weighted Video Quality Metric}
\cref{fig:flowchart} gives an overview of the processing steps of our metric PW-VQM.
\begin{figure}[t]
    \centering
    	\begin{tikzpicture}[
		dataBlob/.style={rectangle, anchor = center, align=center},
		Layer/.style={rectangle, draw, anchor=center, minimum width=5.1cm, minimum height=.55cm, rounded corners, align=center},
		Misc/.style={rectangle, draw, anchor = center, fill=black!10, minimum height=0.75cm},
		Code/.style={, draw, anchor = center, fill=black!20},
		bg/.style={rectangle, anchor = center, fill=\colorBG, anchor = north west,},
		myArrow/.style={-stealth},
		label/.style={rectangle, anchor = center},
		operator/.style={circle ,draw, anchor = center, minimum size = 0.2cm, inner sep = 0pt},
		dot/.style={circle, fill, minimum size=3pt, inner sep=0pt}
		]
        \tikzstyle{every node}=[font=\footnotesize]

        \node (input) [dataBlob] {Luminance Reference\\+ Distorted Frames};
        \node (input2) [dataBlob, right of=input, node distance=4cm] {Reference Video\\Frames (RGB)};
        \node (step1) [Layer, below of=input, node distance=.85cm] {Contrast Sensitivity Function};
        \node (step2) [Layer, below of=step1, node distance=.75cm] {Multiscale Wavelet Transform};
        \node (step3) [Layer, below of=step2, node distance=.75cm] {Structural Similarity Map};
        \node (step4) [Layer, below of=step3, node distance=.75cm] {Weighted Subband Merging};
        \node (step5) [Layer, below of=step4, node distance=.75cm] {Perceptually Weighted Averaging};
        \node (optical_flow) [Layer, right of=step3, node distance=4cm, yshift=.23cm, densely dashed, minimum width=1.7cm] {Optical\\Flow\\Analysis};
        \node (detection) [Layer, below of=optical_flow, node distance=1.05cm, densely dashed, minimum width=1.7cm] {Object\\Detection};
        \node (output) [dataBlob, below of=step5, node distance=.75cm] {Framewise PW-VQM $\in [0, 100]$};

        \node (foreground_label) [dataBlob, above of=optical_flow, node distance=1.05cm] {Foreground\\Detection};

        \begin{pgfonlayer}{background}
			\node (foreground) [rounded corners, draw, fit=(optical_flow)(detection)(foreground_label), inner sep=3pt] {};
		\end{pgfonlayer}

        \draw[myArrow] (input) -- (step1);
        \draw[myArrow] (step1) -- (step2);
        \draw[myArrow] (step2) -- (step3);
        \draw[myArrow] (step3) -- (step4);
        \draw[myArrow] (step4) -- (step5);
        \draw[myArrow] (step5) -- (output);
        \draw[myArrow] (input2) -- (foreground);
        \draw[myArrow] (foreground) |- (step5);

	\end{tikzpicture}
    \caption{Flowchart of the processing pipeline to obtain the frame-wise PW-VQM.}
    \label{fig:flowchart}
    \vspace{-10pt}
\end{figure}
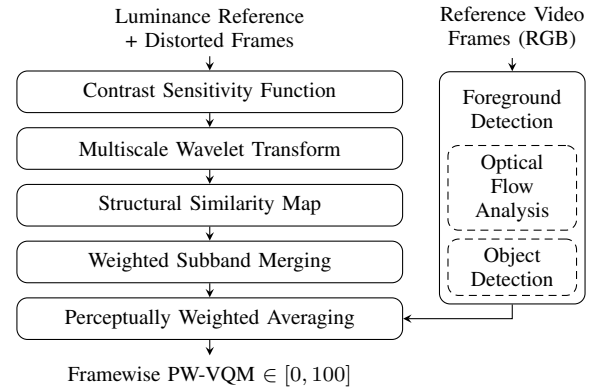
\begin{figure*}[tb]
	\centering 
    	\subfloat[Reference frame]{
	\includegraphics[width=0.24\textwidth]{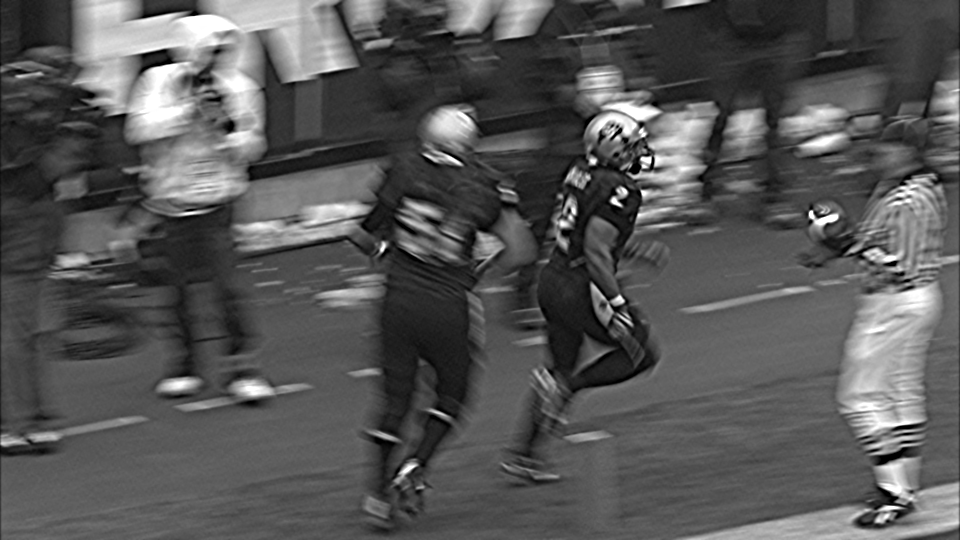}}
    	\subfloat[Object detection map]{
	\includegraphics[width=0.24\textwidth]{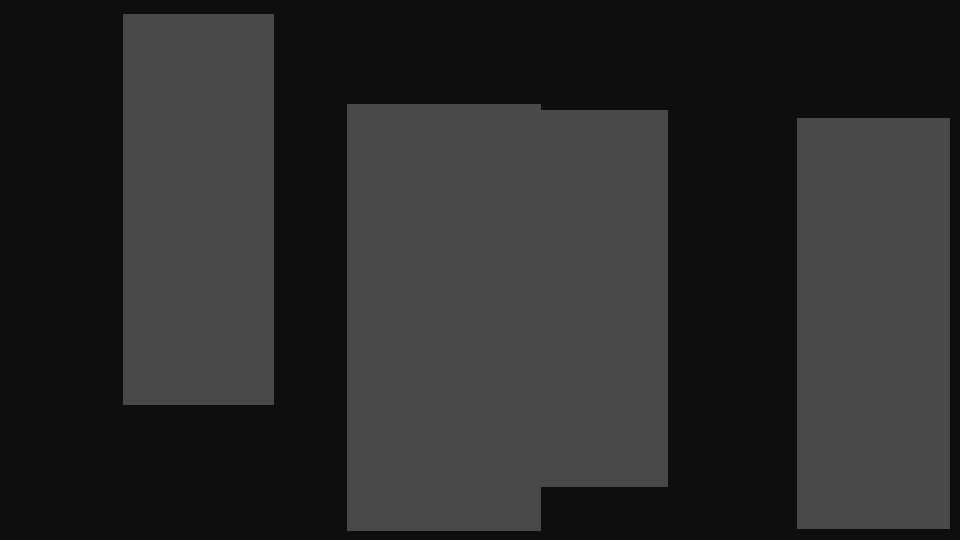}}
          	\subfloat[Object detection + flow map]{
	\includegraphics[width=0.24\textwidth]{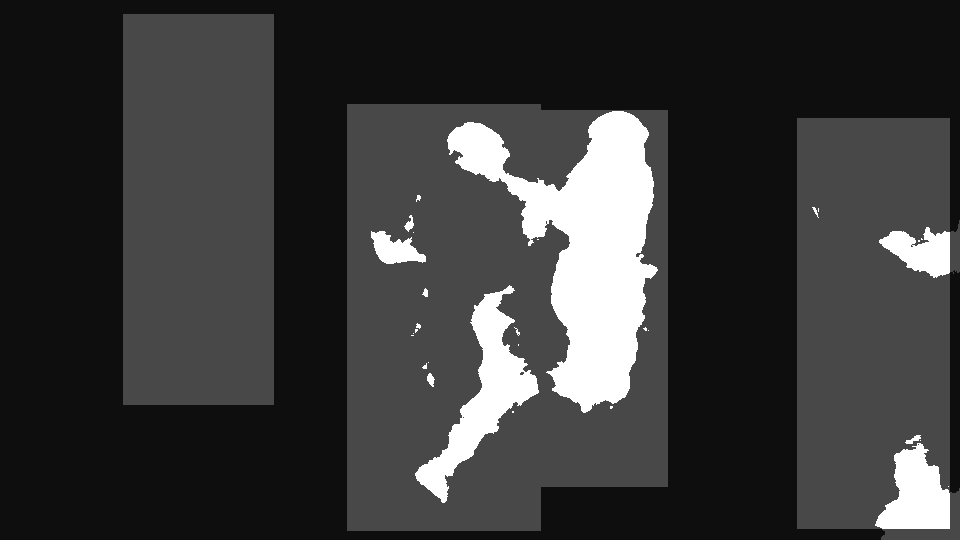}}
	\caption{Visualization of the foreground maps for \textit{football\_1080p\_002\_base\_QP\_15\_G1\_QP\_15\_G2\_QP\_15\_res\_960x540} from the Sports-ROI data set. Both the source and distorted frame are provided for reference. For the two foreground maps in (c)-(d), black corresponds to background. The brighter the region, the more important it is for perception based on our two approaches. The football players are detected by the bounding boxes in (c). The additional flow map in (d) identifies the players in the foreground to be important for the viewer, as they are less affected by motion blur. }
    \label{fig:vis}
     \vspace{-10pt}
\end{figure*}
\\
\textbf{1. Contrast Sensitivity Function (CSF):}
We apply the Fourier-based contrast sensitivity model of Ngan~\emph{et.~al.}~\cite{ngan1989} on the luminance values of the input frames.
For efficient computation, we follow the spatial domain filter approach to avoid the calculation of Fourier transforms~\cite{venkataramanan2024}.
The filter size is set to $N=21$, with viewing distance $\frac{D}{H}=3.5$.
\\
\textbf{2. Multiscale Wavelet Transform:}
The input is transformed into a multiscale wavelet-based subband representation using the coif5 wavelet transform from the Coiflet family~\cite{daubechies1992}.
We set the number of scales to $S=3$, so each frame is decomposed into 10 subbands.
\\
\textbf{3. Structural Similarity Map:}
For each subband, we derive the SSIM map~\cite{wang2004} using a block window of size $8\times8$.
\\
\textbf{4. Weighted Subband Merging:}
The multiscale subband maps are weighted according to Mannos and Sakrison~\cite{mannos1974}, resized to the frame resolution by nearest neighbor upsampling, and merged into one map via a summation across the different subband maps. For weighting, the lowest frequency of the respective subband is chosen. Additionally, the frequency of the diagonal band is weighted less, according to \cite{larson2010}, to account for the oblique effect \cite{dal1992}. Overall, we obtain a single full-resolution similarity map. 
\\
\textbf{5. Foreground Detection:}
In order to distinguish between the perceptually more important foreground and the background, we combine two methodologies based on the reference sequence.
First, we obtain \textbf{object bounding boxes} by using Grounding DINO~\cite{liu2024}.
Since we are working with sports content, we detect objects of the classes "player", "referee", and "sports ball" with a logit threshold of~$0.3$ for each case.
However, to make the detection of the sports ball more robust, we only detect a single ball per frame with the highest logit.
We also linearly interpolate the bounding boxes of the sports ball if no ball was detected, but the previous and next frame contain consistent detections.
All detected bounding boxes are considered to be perceptually more important foreground.
Second, we analyze the \textbf{optical flow} field calculated between the current frame and the previous frame using SPyNet~\cite{ranjan2017}. We consider regions of large local flow as background because we assume that the image content in those regions moves too fast for a human viewer to perceive fine details. To obtain a foreground flow map, we normalize optical values to frame resolution and consider values smaller than~$40$ to represent foreground. Note that this reasoning only holds for very large motion, where the frame content can be affected by distortions like motion blur, whereas human attention is drawn to slower-moving regions of a frame. The two methodologies described above segment the current frame into perceptually important foreground and background regions. The background regions have a value of one, while the foreground regions are assigned a fixed foreground weight~$>1$. We combine the two foreground segmentation maps by multiplication to extend the overall weight range and direct the focus of our metric more towards important frame regions. \cref{fig:vis} provides a visualization of the foreground detection maps for a sample sports video.
\\
\textbf{6. Perceptually-Weighted Averaging:}
We combine both the merged similarity map from step 4. with the foreground detection from step 5., by calculating a weighted average of the similarity map.
We use a foreground weight value of~$5$.

Through our frame-wise processing, we obtain per-frame quality estimates.
In order to aggregate a single quality metric per video sequence, we take the mean of the frame-wise PW-VQM values.
Optionally, the frame-wise PW-VQM metric can only be calculated for every $n$-th frame to reduce computational complexity.

\section{Experimental Results}

\subsection{Settings}
We evaluate our metric PW-VQM on the asymmetric encoded Sports-ROI data set made available through the QoMEX 2026 Grand Challenge on \textit{Video Quality Assessment for Asymmetric Encoded Videos}. The publicly available challenge data sets consists of 155 1080p video sequences with high motion team sports video content, including basketball, football, and soccer. The sequences are distorted by scaling artifacts as well as H.265/HEVC compression. Next to symmetrically encoded content, where all regions are encoded by the same Quantization Parameter (QP), the data set contains asymmetrically coded videos. Here, different QP offsets were applied in regions of high and low importance. Important regions were determined by predicted visual saliency or semantically important regions, i.e., players, field lines, balls, and referees. The subjective quality scores were collected on a 4K display in a side-by-side comparison of reference and distorted video.

Secondly, we consider an additional symmetric encoded sports database, because no further asymmetric encoded data sets with human ratings are available. The LIVE Livestream data set \cite{shang2021a} \cite{shang2021b} consists of 45 source sequences in 1080p or 4K resolution with high motion sports content. The data set contains 10 different types of sports, including team sports as in the Sports-ROI data set, and individual disciplines like running or skateboarding. We perform our experiments on sequences distorted by H.264/AVC compression, representing a symmetric encoded sports data set. The subjective quality scores were collected from 40 subjects.

Our metric PW-VQM does not require training, as we use pretrained models for optical flow estimation and object detection\footnote{For Grounding DINO, we use the \textit{GroundingDINO-B} checkpoint, and for SpyNet, the \textit{sintel-final} model.}. Still, we designed and optimized different components of our metric for the asymmetrically coded Sports-ROI challenge data set, as will be discussed in the ablation study in \cref{sec:ablation}.

\subsection{Results}
\begin{table}[tb]
    \centering
    \caption{Results of Spearman's Rank Order Correlation Coefficient (SROCC) and Pearson's Linear Correlation Coefficient (PLCC) on different data sets and quality metrics. A higher value indicates better alignment with human judgment. 
    \textbf{Bold} and \underline{underlined} values indicate the best and second best result.}
    \begin{tabular}{c|cc|cc}
         & \multicolumn{2}{c}{Sports-ROI} &  \multicolumn{2}{|c}{LIVE} \\
         Metric & SROCC $\uparrow$ & PLCC $\uparrow$  & SROCC $\uparrow$ & PLCC $\uparrow$ \\
         \midrule
         PSNR-Y                             & $0.6940$ & $0.6732$ & $0.7909$ & $0.9355$ \\
         SSIM~\cite{wang2004}               & $\underline{0.9038}$ & $0.9235$ & $\underline{0.9224}$ & $\textbf{0.9745}$ \\
         MS-SSIM~\cite{wang2003}            & $0.8865$ & $0.8972$ & $0.8920$ & $0.9517$ \\
         VMAF                               & $0.8225$ & $0.9123$ & $0.8613$ & $0.9523$ \\
         VMAF\_NEG                          & $0.8520$ & $0.9258$ &  $0.8709$ & $0.9534$ \\
         VMAF\_4k                           & $0.8682$ & $0.9231$ & $0.8672$ & $0.9523$ \\
         VMAF\_4K\_NEG                      & $0.8787$ & $\underline{0.9353}$ & $0.8656$ & $0.9539$ \\
         Funque \cite{venkataramanan2022}   & $0.7397$ & $0.8686$ & $0.8622$ & $0.9402$ \\
         LPIPS \cite{zhang2018}             & $0.8912$ & $0.9019$ & $0.8228$ & $0.9064$ \\
         \textbf{PW-VQM (ours)}             & $\textbf{0.9511}$ & $\textbf{0.9573}$  & $\textbf{0.9235}$ & $\underline{0.9697}$ \\
    \end{tabular}
    \label{tab:results}
        \vspace{-10pt}
\end{table}
\cref{tab:results} shows the results of the proposed metric compared to different common video quality metrics. The metrics used for comparison are PSNR-Y, SSIM~\cite{wang2004}, MS-SSIM~\cite{wang2003}, Funque \cite{venkataramanan2022}, LPIPS \cite{zhang2018}, and different configurations of VMAF. We use the implementation available through libvmaf or their respective official repository.

The correlation of the metrics to the subjective quality scores is measured by Spearman's Rank Order Correlation Coefficient (SROCC) and Pearson's Linear Correlation Coefficient (PLCC).
Before the calculation of PLCC, a five-parameter logistic function is fitted to the quality scores according to \cite{shang2012}: 
\begin{equation}
     \hat{q}(q) = \beta_1 \left(\frac{1}{2} - \frac{1}{1+ \text{e}^{\beta_2 (q-\beta_3)}}\right) + \beta_4 \cdot q + \beta_5.
\end{equation}
Thereby, $q$ and $\hat{q}$ denote the unfitted and fitted quality scores of the metrics, respectively. The parameters $\beta_1, \beta_2, \beta_3, \beta_4$ and $\beta_5$ are fitted for each metric individually to the subjective Mean Opinion Scores (MOS) of the data set using nonlinear least squares fitting. 

As can be seen in \cref{tab:results}, our proposed metric performs best on the Sports-ROI data set both in terms of SROCC and PLCC. VMAF\_4K\_NEG is second-best in terms of PLCC, while SSIM is second-best in terms of SROCC. With an SROCC score of 0.9511, our metric significantly outperforms SSIM with an SROCC score of 0.9038,  demonstrating its effectiveness for asymmetric encoded sports content. For symmetric encoded sports content, i.e., the LIVE data set, our metric PW-VQM performs close to SSIM but significantly better than VMAF and its variants. This demonstrates that the sports-specific prompts used for the object detection weighting generalize to other data sets with different characteristics and sports activities.

In the full reference track of the QoMEX 2026 grand challenge, our metric PW-VQM ranked third out of 8 teams for the asymmetric encoded non-public test data set. According to the initial results of the challenge, the best team achieved an SROCC score of 0.877, the second best a score of 0.864, and our metric scored 0.839, which indicates that our method is promising for asymmetric encoded sports content.

\subsection{Ablation Study} 
\label{sec:ablation}
To verify the effectiveness of the proposed perceptual weighting for sports content, we perform an ablation study on the Sports-ROI data set. \cref{tab:ablationstudy} contains SROCC scores for different configurations of our metric. Without modeling contrast sensitivity, we obtain an SROCC score of $0.8337$. Enabling the spatial CSF gives an improved value of $0.9355$. Additionally, considering contrast sensitivity in the wavelet domain improves the SROCC score to $0.9383$. When spatial weighting is further enabled, our results demonstrate that both the object detection and flow weighting individually lead to a higher SROCC score. We optimized the foreground weight value for object detection weighting only, where a foreground weight of 5 performed best. The flow weighting threshold of 40 was also determined on the Sports-ROI data set. Finally, the results in \cref{tab:ablationstudy} show that combining object detection and flow weighting results in the overall best SROCC score of $0.9511$.

\begin{table}[t]
\centering
\caption{Ablation study on the Sports-ROI data set.}
\begin{tabular}{cccc | c}
\toprule
Spatial CSF & Wavelet CSF & Object detection&  Flow & SROCC $\uparrow$ \\
\midrule
\xmark  & \xmark  & \xmark  & \xmark  & $0.8337$ \\
\cmark & \xmark  & \xmark  &  \xmark & $0.9355$ \\
\cmark & \cmark & \xmark  & \xmark  & $0.9383$ \\
  \cmark&  \cmark& \xmark  & \cmark & $0.9421$ \\ 
  \cmark  & \cmark &  \cmark& \xmark  & $0.9483$ \\ 
    \cmark    & \cmark & \cmark & \cmark & $0.9511$ \\ 
\bottomrule
\end{tabular}
\label{tab:ablationstudy}
    \vspace{-10pt}
\end{table}

\section{Conclusion}
In this paper, we proposed a metric for evaluating the visual quality of sports content by applying perceptual weighting on relevant video regions.
Especially for asymmetrically encoded content, we significantly outperform a baseline of common video quality metrics, while we also perform best for symmetrically encoded sports content in terms of SROCC.
In future work, the metric can be extended to support different types of asymmetric encoded content. In addition, the computation time can be analyzed and optimized.

\end{document}